%% file: main.tex
\documentclass[11pt]{article}

\usepackage{tikz}
\usepackage{multirow}
\usepackage{enumitem}
\usetikzlibrary{arrows.meta, positioning}
\usepackage{amsmath, amssymb, amsthm}
\usepackage{graphicx}
\usepackage{booktabs}
\usepackage{hyperref}
\usepackage{fullpage}
\usepackage[numbers]{natbib}

\usepackage{parskip}

\usepackage{xcolor}
\newcounter{todocounter}
\begin{document}

\title{Systematic Review of Academic Procrastination Interventions in Computing Higher Education}

\author{
Daniel Cheng \\
University of Toronto \\
daniel.cheng@mail.utoronto.ca
\and 
Oscar Heath \\
University of Toronto \\
oscar.heath@mail.utoronto.ca
\and
Daniyaal Farooqi \\
University of Toronto \\
daniyaal.farooqi@mail.utoronto.ca
\and
Evelyn Chou \\
Stanford University \\
evelyn.chou@mail.utoronto.ca
\and
Alice Gao \\
University of Toronto \\
ax.gao@utoronto.ca
\and
Jonathan Calver \\
University of Toronto \\
calver@cs.toronto.edu
}

\maketitle

\begin{abstract}
Academic procrastination is a persistent challenge in computing education, yet evidence on the effectiveness of course-level interventions remains fragmented across diverse designs and contexts. We present a systematic literature review of studies published in the past decade that empirically examine interventions to reduce academic procrastination among post-secondary computing students. Evidence from 19 articles examines interventions that target procrastination through structural, feedback-based, motivational, and self-regulatory mechanisms.
Our findings suggest that interventions introducing clear temporal structure consistently promote earlier starts and more distributed work, which act as key mediators of performance gains. The magnitude of these gains depends strongly on task structure, with greater benefits for long-horizon, multi-step assignments than for short, routine tasks. Moreover, supportive designs reliably outperform punitive or restrictive schemes, while uniform interventions yield uneven benefits across students.
This review highlights the importance of designing structured, supportive, and personalized interventions to address procrastination in computing education.
\end{abstract}

\input{1_introduction}

\input{2_related_work}

\input{3_methods}

\input{figure_prisma}

\section{Results}
\label{sec:results}
\input{results_participants}
\input{figure_papers_table}
\input{results_1_deadlines}

\input{results_2_autograding}

\input{results_3_gamification}

\input{results_5_reminders}
\input{results_6_others}

\input{4_discussion}

\input{5_limitations}
\input{6_conclusion}

\section*{Acknowledgements}

We thank Sadia Sharmin for their feedback on earlier drafts of this paper.

\bibliographystyle{unsrtnat}
\bibliography{interventions}

\end{document}

%% file: 1_introduction.tex
\section{Introduction}

Procrastination has been defined as the intentional delay of tasks despite knowing the potential negative consequences \cite{steel2007nature}. Academic procrastination is a pervasive issue affecting students across various disciplines, significantly hindering academic performance and personal well-being. 

Students in computing programs face a distinct set of challenges that contribute to academic procrastination. For instance, many students postpone starting their work because a complete software development cycle, encompassing analysis, design, coding, testing, and documentation, can be unfamiliar and substantially more time-consuming than coursework in many other disciplines \cite{beaubouef2005why}. Beyond course structure, students often face significant mental health challenges, including high levels of stress, anxiety, and depression, which can further exacerbate academic procrastination \cite{sirois2013procrastination}. In fact, attrition rates in computer science are reported to be as high as 30–40\% at many institutions, with most withdrawals occurring during the first two years of study \cite{beaubouef2005why}. When asked about their own study habits, students in computer information systems courses reported that they would produce higher-quality work and be better students if they procrastinated less \cite{schultz2017procrastination}. Consistent with these perceptions, several studies show that students who begin coding assignments earlier tend to produce more accurate programs and earn higher grades \cite{kazerouni2017quantifying, edwards2009comparing}.


It follows that interventions should be tailored to address the unique factors contributing to academic procrastination among computing students. Moreover, the type of coursework in computing programs also provides a unique opportunity to implement tailored interventions, such as automated feedback on a coding assignment \cite{denny2021promoting}. To understand how such interventions have evolved over the past decade, we conduct a systematic literature review on intervention studies designed to reduce procrastination in computing contexts, examining their methodologies and effectiveness. Our review addresses the following research questions:

\begin{itemize}[leftmargin=0.4cm]
    \item \textbf{RQ1}: What interventions have been evaluated to combat academic procrastination in computing courses or among computing students at the post-secondary level?
    \item \textbf{RQ2}: What are the effects of these interventions on student behaviour and course outcomes?
\end{itemize}

After identifying gaps in existing reviews and outlining our methodology, we address the research questions by categorizing interventions and synthesizing patterns within and across categories.

    

%% file: 2_related_work.tex
\section{Related Work}

Research on academic procrastination has expanded rapidly over the past three decades, leading to a growing number of review and synthesis efforts \cite{tao2021bibliometric}. These reviews vary widely in scope, emphasizing correlates, theoretical perspectives, or intervention strategies. 

Several reviews focus on factors associated with general or academic procrastination, such as motivation, metacognition, and self-regulatory skills \cite{hussin2023systematic, feyzi2022exploring}. While these works offer valuable insight into why procrastination occurs, they do not examine how specific instructional interventions are empirically evaluated. 

Beyond correlational analysis, other reviews emphasize the theoretical foundations underlying procrastination interventions \cite{salguero2023interventions, furlan2022interventions}. For example, one review \cite{salguero2023interventions} mapped interventions to psychological dimensions, while another \cite{furlan2022interventions} organized interventions by theoretical frameworks and program characteristics across diverse contexts. However, these reviews place less emphasis on synthesizing empirical evaluations within specific instructional settings.

A third line of work reviews intervention effectiveness across heterogeneous populations.  Several reviews examine both general and academic procrastination among mixed samples of students and adults \cite{rozental2018targeting, van2018overcoming, pereira2025investigating}, whereas our review focuses on academic procrastination in post-secondary computing education. These syntheses span diverse intervention types (including psychological and gamification-based interventions) and various educational contexts, limiting their ability to speak to computing-specific approaches. 

Similarly, \citet{turner2023systematic} reviewed controlled experiments on academic procrastination interventions across multiple educational levels, from primary to tertiary education, rather than focusing exclusively on post-secondary contexts as in our review. Further, this work \cite{turner2023systematic} does not examine how these approaches are applied within computing education.

Taken together, prior reviews emphasize correlates, theoretical perspectives, or interventions evaluated across diverse populations and settings, providing limited synthesis of empirical evidence on intervention effectiveness within specific instructional contexts. This review addresses this gap by synthesizing empirical studies of interventions targeting academic procrastination in post-secondary computing education.

%% file: 3_methods.tex
\section{Methods}
\label{sec:methods}

Our systematic literature review followed the PRISMA 2020 guidelines \cite{page2021prisma} to ensure transparent and reproducible reporting. We searched six databases: ACM Digital Library, IEEE Xplore, ScienceDirect, Scopus, SpringerLink, and Web of Science. The search included research articles published in English between January 1, 2015 to June 30, 2025 with full text available. Studies were eligible if they empirically evaluated an intervention designed to reduce academic procrastination in a post-secondary computing course or among computing students.

Search queries were adapted to the syntax and capabilities of each database. We required that equivalent forms of both ``computing'' and ``procrastination'' appear in at least one of the title, abstract, or keywords fields. For most databases (IEEE, Web of Science, Scopus, and Springerlink), we used the wildcard terms \verb+comput*+ and \verb+procrast*+. Because ACM primarily contains computing-related articles, we searched for \verb+procrast*+ only. ScienceDirect does not support wildcards, so we used explicit variants: compute, computing, computer, and procrastinate, procrastinating, procrastination.

Figure~\ref{prisma_diagram} summarizes the paper selection process. All included papers were analyzed using an iterative, consensus-based qualitative synthesis conducted by a team of five researchers. Each paper was independently reviewed by two researchers, who summarized the intervention design and reported outcomes. Studies were then grouped by intervention category (deadlines, auto-grading, gamification, reminders, psychological, and social), with the full team refining these categories until consensus was reached. For each intervention category, one researcher synthesized key findings describing recurring patterns in intervention effects, which were reviewed by at least one additional researcher. The team met regularly to resolve disagreements and identify patterns spanning multiple intervention categories.

%% file: figure_prisma.tex
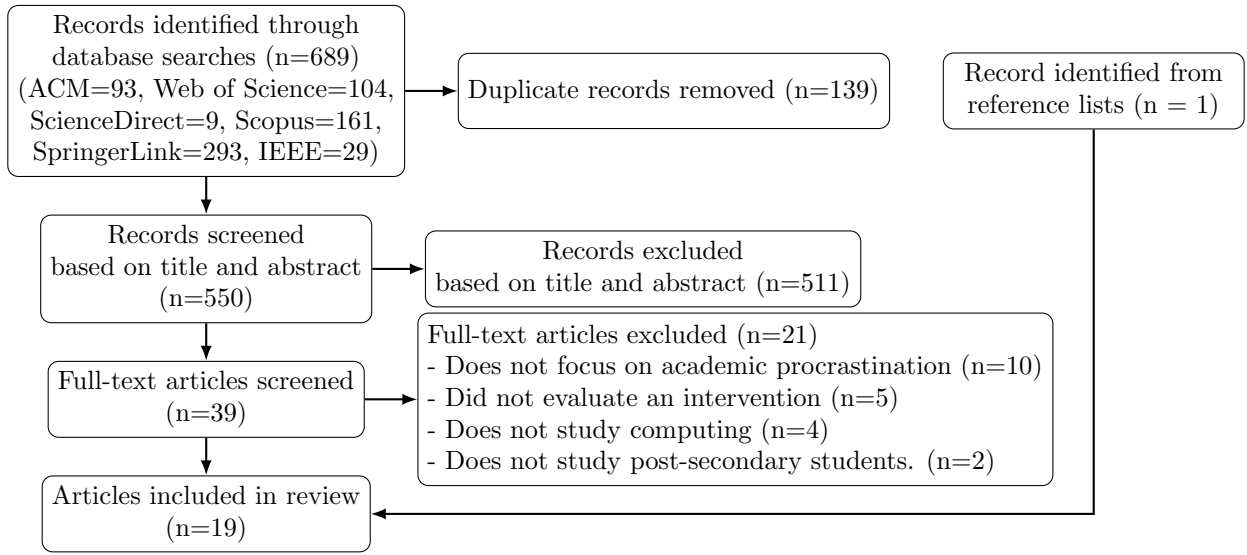
\begin{figure}[ht]
\centering
\begin{tikzpicture}[
  font=\small,
  node distance=5mm and 7mm,
  process/.style={rectangle, rounded corners, draw,
    align=center, minimum width=4cm, minimum height=10mm, fill=white},
  exclude/.style={rectangle, rounded corners, draw, align=center,
                  minimum width=4cm, minimum height=10mm, fill=white},
  arrow/.style={-{Latex[length=2mm]}, thick}
]

\node (n1) [process]
  {Records identified through \\
  database searches {(n=689)}\\
  (ACM=93, Web of Science=104, \\
  ScienceDirect=9, Scopus=161, \\
  SpringerLink=293, IEEE=29)};


\node (n3) [process, below=of n1]
  {Records screened \\
  based on title and abstract\\
  {(n=550)}};

\node (n4) [process, below=of n3]
  {Full-text articles screened\\
  {(n=39)}};

\node (n5) [process, below=of n4]
  {Articles included in review\\
  {(n=19)}};

\node (e2) [exclude, right=of n1]
  {Duplicate records removed (n=139)};

\node (e3) [exclude, right=of n3]
  {Records excluded \\
  based on title and abstract (n=511)};

\node (e4) [exclude, right=of n4, align=left]
  {Full-text articles excluded {(n=21)}\\
   - Does not focus on academic procrastination {(n=10)}\\
   - Did not evaluate an intervention {(n=5)}\\
   - Does not study computing {(n=4)}\\
   - Does not study post-secondary students. {(n=2)}};

\node (a1) [process, right=of e2]
    {{Record identified from} \\
    {reference lists (n = 1)}};

\draw[arrow] (n1) -- (n3);
\draw[arrow] (n3) -- (n4);
\draw[arrow] (n4) -- (n5);

\draw[arrow] (n1.east) -- (e2.west);
\draw[arrow] (n3.east) -- (e3.west);
\draw[arrow] (n4.east) -- (e4.west);

\draw[arrow] (a1.south) |- (n5.east);

\end{tikzpicture}

\caption{PRISMA diagram: paper selection procedure}
\label{prisma_diagram}
\end{figure}

%% file: results_participants.tex
Among the 19 studies in our review, 
12 were performed in North America \cite{she2024clearmind, ye2022behavioral, bernuy2021investigating, yeckehzaare2023reducing, shaffer2021impact, martin2015effects, edwards2015examining, chiu2024Effect, irwin2019can, yeckehzaare2019spaced, bouvier2021overnight, brown2021nudging}, with nine based in the United States \cite{yeckehzaare2023reducing, shaffer2021impact, martin2015effects, edwards2015examining, chiu2024Effect, irwin2019can, yeckehzaare2019spaced, bouvier2021overnight, brown2021nudging}. Five were in Europe \cite{pereira2021academic, auvinen2015increasing, ibanez2019using, castro22experiences, hakulinen2015effect} and two were in Australasia \cite{leinonen2022comparison, denny2021promoting}.
Furthermore, all 19 studies focused on undergraduate students, and only one included graduate students~\cite{yeckehzaare2023reducing}.
In terms of course context, ten studies focused on introductory or post-introductory programming courses \cite{ye2022behavioral,bernuy2021investigating, irwin2019can,bouvier2021overnight,brown2021nudging,yeckehzaare2019spaced,leinonen2022comparison,denny2021promoting,ibanez2019using,chiu2024Effect},
five examined data structure and algorithm courses \cite{auvinen2015increasing, edwards2015examining, hakulinen2015effect, martin2015effects, shaffer2021impact}, while the remaining papers studied data science courses \cite{she2024clearmind},
three courses on CS and statistics \cite{castro22experiences}, 
a software engineering capstone project \cite{pereira2021academic}, 
and an independent studies course on web development \cite{yeckehzaare2023reducing}.


Table~\ref{tab:included_papers} organizes the included interventions by category. Each category section begins with a summary of the interventions, followed by an interpretative synthesis of the key findings that highlights recurring patterns within that category. Finally, Section \ref{sec:discussion} integrates the findings to identify broader patterns across categories.

%% file: figure_papers_table.tex
\begin{table}[ht]
\centering
\begin{tabular}{p{0.15\linewidth}p{0.6\linewidth}p{0.13\linewidth}}

\toprule
\textbf{Category} &
\textbf{Intervention} &
\textbf{References} \\

\midrule
\multirow{3}{*}{Deadlines} 
& Various deadline types (none, suggested, soft, hard) & 
\cite{chiu2024Effect} \\
& Interim deadlines in 3-4 week project & 
\cite{shaffer2021impact} \\
& Deadlines with varying days and times & \cite{castro22experiences} \\

\midrule
\multirow{4}{*}{Auto-grading} 
& Scheduled feedback at 10am daily & \cite{bouvier2021overnight} \\
& Scheduled feedback at 2 optional deadlines & \cite{denny2021promoting, leinonen2022comparison} \\
& Limited feedback using regenerating tokens & \cite{irwin2019can} \\
& Unlimited feedback w/ penalty after 2nd attempt & \cite{leinonen2022comparison} \\

\midrule
\multirow{5}{*}{Gamification} 
& Heatmap visualizations of progress vs. class & \cite{auvinen2015increasing} \\
& Achievement badges rewarding time management & \cite{hakulinen2015effect} \\
& Augmented reality game vs. non-gamified quiz & \cite{ibanez2019using} \\
& Limited feedback using regenerating tokens & \cite{irwin2019can} \\
& Practice tool w/ daily goals, progress tracking, \& celebratory fireworks & \cite{yeckehzaare2019spaced} \\

\midrule
\multirow{5}{*}{Reminders} 
& Email reminders with personalized feedback& \cite{martin2015effects, edwards2015examining} \\
& Email reminders with varying send time & \cite{bernuy2021investigating} \\
& GitHub issue reminders & \cite{brown2021nudging} \\
& Email reminders with varying content & \cite{ye2022behavioral} \\
& CBT-based chatbot for self-regulation & \cite{pereira2022struggling} \\

\midrule
\multirow{2}{*}{Psychological} 
& Voluntary weekly presentations to discuss work & \cite{yeckehzaare2023reducing} \\
& ACT workshop on procrastination & \cite{she2024clearmind} \\
or Social & Reflective written assignment, schedule sheets & \cite{martin2015effects, edwards2015examining} \\
\bottomrule
\end{tabular}
\caption{Included papers categorized by intervention type}
\label{tab:included_papers}
\end{table}

%% file: results_1_deadlines.tex
\subsection{Deadlines}

Structured deadlines are a common instructional tool used to organize coursework and guide students' efforts. By varying deadline placement and late penalties, educators aim to encourage timely progress and improve course outcomes.
Three papers \cite{chiu2024Effect, shaffer2021impact, castro22experiences} studied the effects of deadlines on procrastination. 
\citet{chiu2024Effect} evaluated multiple deadline policies with increasing structure and accountability (no, suggested, soft, and hard deadlines) in an online self-paced programming course. Suggested deadlines had no grade penalty for late submissions, whereas soft or hard deadlines had minor or major late penalties respectively.
\citet{shaffer2021impact} measured the effect of adding three interim deadlines, called milestones, in a programming project spanning 3--4 weeks. The milestones were evaluated by auto-tests, worth at most 10\% of the project grade.
Finally, \citet{castro22experiences} studied the effects of the day and time of weekly deadlines on student submission behaviour. They evaluated six deadline placements varying the day of the week and the time of the day (afternoon, evening or midnight) 
in three courses on programming, software development, and statistics.
%

\emph{Deadlines substantially reduce late submissions and improve submission timeliness.}
In a self-paced online programming course, 
all deadline policies, including deadlines with no penalties, substantially reduced late submissions compared to having no deadlines \cite{chiu2024Effect}. More structured deadlines were especially effective, with a hard mid-term deadline producing the strongest improvements in submission timeliness \cite{chiu2024Effect}. Similarly, creating milestones in large programming projects encouraged earlier submissions \cite{shaffer2021impact}. Within the treatment group, students who completed more milestones submitted substantially earlier than the rest \cite{shaffer2021impact}. 

\emph{Deadline placement affects submission timing and concentration.}
\citet{castro22experiences} found that students tended to work closer, rather than further away, from the deadlines.
Deadlines later in the week led to fewer submissions near the deadlines compared to deadlines earlier in the week, likely due to students working over the weekend. Specifically, students submitted closest to the deadline at 4 PM on Monday and farthest in advance for a deadline on Friday at midnight.
Deadline time also mattered: midnight deadlines produced smoother submission patterns with fewer last-minute spikes. Daytime deadlines (4 PM or 6 PM) showed sharp surges in activity immediately before the deadline.
To summarize, deadlines later in the week and during night time can effectively discourage last minute student work compared to deadlines earlier in the week and during day time.

\emph{Deadlines meaningfully affect student performance and course outcomes.} 
\citet{chiu2024Effect} found that in a self-paced programming course, changes to deadline structure led to measurable differences in pass, incomplete, and withdrawal rates \cite{chiu2024Effect}. In contrast, intermediate deadlines in multi-week programming projects primarily affected academic performance, improving correctness scores and final grades for mid-performing students without altering pass or withdrawal rates \cite{shaffer2021impact}. Beyond overall structure, deadline placement also shaped performance, as submissions made closer to deadlines and those completed during night time showed lower correctness than earlier or day-time submissions \cite{castro22experiences}. Overall, deadline design influences student outcomes through multiple pathways, with overall structure shaping retention and completion, while intermediate deadlines and deadline placement primarily affect timing and quality of students' work.

\emph{Supportive deadlines improved course outcomes and reduced stress compared to punitive ones.}
In a self-paced programming course, supportive deadlines with no grade penalty achieved better outcomes (higher pass rates, fewer incompletes and withdrawals) than deadlines with even a minor grade penalty \citep{chiu2024Effect}. Penalty-based deadlines appeared to increase stress, and dropping the lowest grade only partially mitigated these negative effects \citep{chiu2024Effect}.
Additional evidence highlights the benefits of supportive deadlines more broadly. In a multi-week project, intermediate milestones improved project correctness and final grades, and most students perceived these milestones as helpful rather than burdensome \cite{shaffer2021impact}.

%% file: results_2_autograding.tex
\subsection{Auto-grading}

Autograders are a distinctive feature of computing education, providing both scalable grading and near-instant feedback to students. However, over-reliance on automated feedback can hinder learning, so instructors often limit students' access to it.
Four papers \cite{denny2021promoting, leinonen2022comparison, bouvier2021overnight, irwin2019can} examine two types of autograder feedback policies aimed at reducing procrastination. 
The first type of policy introduces non-graded early deadlines in a programming project where students receive automated feedback.
\citet{denny2021promoting} evaluated a 20-day programming project with a ``scheduled feedback'' policy, having two early deadlines at two weeks and one week before the final deadline. Students who submitted by each deadline received minimal automated feedback reporting the percentage of test cases passed per task, without revealing specific errors or solutions.
\citet{bouvier2021overnight} studied an ``overnight feedback'' policy in an 8-day programming project. Students who submitted by 10 PM received automated feedback by 10 AM the following morning, consisting of compilation results and program output for predefined test cases.
The second type of policy provides on-demand feedback with usage constraints.
\citet{irwin2019can} introduced ``submission energy'' inspired by mobile games, evaluated on programming assignments spanning two weeks. Each student could have up to three energy units. Each request for automated feedback consumed one unit, which regenerated after one hour. The feedback consisted of automated test results on instructor-defined test cases. 
\citet{leinonen2022comparison} examined ``immediate feedback'', a penalty-based policy in the same 20-day programming project in~\cite{denny2021promoting}. During the final 5 days, students could receive automated test results on demand, showing the proportion of test cases passed. However, each attempt after the first two incurred a 10\% grade penalty up to a maximum of 70\%.

\emph{Policies providing clear temporal signals effectively prompt earlier engagement and reduce lateness.}
Scheduled feedback produced pronounced submission spikes, with 49\% of students submitting before the first early deadline and 70\% before the second \cite{denny2021promoting}.
Similarly, overnight feedback substantially increased submissions made at least one day early (from 14\% to 42\%), while reducing late submissions (from 30\% to 5\%) \cite{bouvier2021overnight}.
Submission energy reduced late submissions significantly (from 14\% to 7\%), though it shifted first submissions only slightly earlier (by 3 hours) \cite{irwin2019can}.
In contrast, immediate feedback with penalties lacked strong temporal signals and did not reliably reduce delayed starts \cite{leinonen2022comparison}.
The results suggest that clear temporal structure around feedback availability matters more for submission timeliness than increasing feedback access alone.

\emph{Structured feedback access can improve work pacing without increasing student workload.}
In \cite{leinonen2022comparison}, both scheduled and immediate feedback encouraged more distributed work rather than last-minute submissions. However, scheduled feedback prompted bursts of activities around the two early deadlines, whereas immediate feedback led to more evenly spaced submissions over time \cite{leinonen2022comparison}. 
Similarly, submission energy promoted more distributed engagement and reduced binge-working, as shown by a larger number of distinct work sessions and fewer assignments completed in a single session \cite{irwin2019can}. However, the time span between first and last submission did not change, indicating redistributed rather than increased total effort \cite{irwin2019can}.

\emph{Earlier engagement, rather than increased feedback access, drives performance gains.}
Delayed first submission was strongly associated with project failure under both scheduled and immediate feedback schemes, despite large differences in feedback availability \cite{leinonen2022comparison}. 
Consistent with this pattern, under scheduled feedback, students who submitted before either early deadline earned substantially higher grades than those who did not \cite{denny2021promoting}.
Most notably, for at-risk students, submitting before interim deadlines under scheduled feedback is associated with substantially higher project scores, whereas greater access to immediate feedback does not yield comparable performance gains \cite{leinonen2022comparison}.
Finally, limiting feedback frequency via submission energy produced significant gains in assignment scores and final grades despite reducing overall feedback access, suggesting that earlier starts and more distributed work may help explain these performance improvements \cite{irwin2019can}. 
Overall, hese results point to engagement timing as the primary pathway through which feedback policies affect performance.

\emph{Students respond more positively to supportive feedback schemes than to those with penalties or constraints.}
Overnight feedback was viewed very positively: many students reported that it was helpful (93\%), improved program quality (88\%), and motivated earlier start (64\%) \cite{bouvier2021overnight}.
In a direct comparison, scheduled feedback was also perceived as more helpful than immediate feedback with penalties. 87.5\% of students agreed that they received helpful feedback under scheduled feedback, compared to 71.7\% under immediate feedback with penalties, despite the latter providing greater feedback availability \cite{leinonen2022comparison}.
Submission energy elicited mixed reactions: while some students (73\%) reported more deliberate submission behavior, many (67\%) expressed frustration with the recharge time \cite{irwin2019can}.
Taken together, these findings indicate that supportive feedback schemes align positive perceptions with their behavioral and performance benefits, whereas punitive or restrictive designs may erode perceived helpfulness despite shaping behaviour.

%% file: results_3_gamification.tex
\subsection{Gamification}

Gamification incorporates game-like design elements into learning activities to influence students’ motivation and engagement. These elements may operate through extrinsic incentives (e.g., points and badges) or by supporting intrinsic motivation (e.g., enjoyment and a sense of progress), making tasks more appealing to begin and sustain.
Five studies \cite{ibanez2019using, yeckehzaare2019spaced, irwin2019can, hakulinen2015effect, auvinen2015increasing} used gamified interventions to combat academic procrastination.
\citet{ibanez2019using} transformed a standard quiz into an augmented reality location-based game, in which students physically traveled to campus locations and answered questions to capture virtual characters.
\citet{yeckehzaare2019spaced} designed a practice tool with gameful features, awarding one point per day for completing a minimum number of questions up to 45 points for the semester. The tool also showed visual progress tracking towards daily/semester goals and celebratory firework animations for daily goal completion. 
As mentioned in the Autograding section, the submission energy in \cite{irwin2019can} draws inspiration from popular mobile game mechanisms. 
This intervention turned automated feedback into a scarce, regenerating resource, encouraging students to start early and spread out their work. 
\citet{hakulinen2015effect} offered three types of achievement badges to improve students' behaviour without affecting grades. The time management badges rewarded completing the tasks early or among the fastest. The carefulness badges rewarded solving the exercises on the first try or with minimal attempts. The learning badges recognized students for earning full points or revisiting exercises after a delay.
\citet{auvinen2015increasing} displayed a heatmap predicting a student’s final points by comparing their current behaviour to that of past students. The visualization incorporated five behavioural variables: number of attempts, first-submission points, interval and improvement between attempts, and earliness of submissions.
%

\emph{Some gamified interventions can promote earlier starts, with varying effects across designs and student populations.}
Three studies show that gamification can motivate students to begin their work earlier. Submission energy \cite{irwin2019can} prompted students to begin their assignments three hours earlier, time-management badges \cite{hakulinen2015effect} motivated students start their work 1.3 days earlier on average, and the daily reward structure \cite{yeckehzaare2019spaced} motivated many students to start using the tool from the first day rather than clustering work near exams.
However, these benefits were not universal: visualizations of peer activity 
did not alter the behaviour of low-performing students \cite{auvinen2015increasing}, 
while an augmented-reality game did not shift start times relative to the control group within the 5-day period \cite{ibanez2019using}.
To summarize, earlier starts emerged primarily when gamification introduced recurring incentives or pacing constraints that rewarded early engagement, rather than relying on novelty or passive feedback.

\emph{Gamification encourages spaced work and reduces deadline pressure.}
%
The submission energy \cite{irwin2019can} resulted in more work sessions per assignment and significantly fewer students completing the assignments in a single sitting, although students' total time spent did not change.
The daily reward structure \cite{yeckehzaare2019spaced} produced substantial student activity on nearly every day of a 45-day period. 
These changes in work patterns were accompanied by earlier task completion and reduced late submissions. The augmented reality game \cite{ibanez2019using} motivated students to finish their tasks before the final day of the 5-day period. Similarly, the submission energy \cite{irwin2019can} and time-management badges \cite{hakulinen2015effect} led to fewer submissions before the deadline and fewer late submissions.
Together, these findings suggest that gamification combats procrastination less by accelerating final effort and more by reshaping students' temporal engagement with coursework.

\emph{Gamification improves engagement and motivates work beyond the minimum requirement.}
Several studies found evidence that gamification motivated students to do more work, sometimes even beyond the explicit requirements, consistent with increased intrinsic motivation rather than compliance with external incentives. The gamified practice tool in \cite{yeckehzaare2023reducing} motivated a third of students to continue using it even after earning the maximum points for the semester. Likewise, students using the augmented reality game \cite{ibanez2019using} answered more questions than those using the non-gamified quiz.
These findings suggest that well-designed gamification can shift students from minimal compliance toward sustained, self-directed engagement.

\emph{Gamification has mixed effects on student performance and course outcomes.}
Some gamified designs were associated with clear performance gains: In \cite{yeckehzaare2019spaced}, each additional hour spent using the gamified retrieval-practice tool was associated with roughly a 1\% increase in final exam scores. In \cite{irwin2019can}, submission energy significantly increased correctness (90.2\% to 93.9\%) and overall scores in the assignments (87.1\% to 91.6\%). 
In contrast, other gamification approaches primarily affected engagement without translating into improved learning outcomes. The augmented-reality quiz game led to earlier task completion but no change in learning outcomes compared to a non-gamified version \cite{ibanez2019using}, and achievement badges led to little or no improvement in overall grades despite positive behavioral changes \cite{hakulinen2015effect}. 
%
%
These results suggest that increased engagement alone is insufficient for improving outcomes, and that gamified designs must align closely with learning processes to yield performance gains.

\emph{Students respond to gamified interventions differently based on performance level and motivational orientation.}
In \cite{auvinen2015increasing}, achievement badges and heatmap visualizations primarily benefited high-\\performing students, leading them to earlier submissions and slightly higher total points, while neither intervention meaningfully engaged low-performing or less self-regulating students.
Moreover, the two designs appealed to different motivational profiles: badges resonated with students motivated by grades and demonstrating strong performance, whereas visualizations attracted students concerned about avoiding poor outcomes \cite{auvinen2015increasing}.
These results highlight the need for personalized gamified designs that account for differences in students’ motivation and self-regulation.

%% file: results_5_reminders.tex
\subsection{Reminders}

Educators have used various forms of reminders as a low-stakes way to reduce procrastination by making deadlines, progress, and next steps more salient in a course.
Six studies \cite{martin2015effects, edwards2015examining, bernuy2021investigating, ye2022behavioral, brown2021nudging, pereira2022struggling} investigated the impact of reminders on student behaviour and performance.
Four studies \cite{martin2015effects, edwards2015examining, bernuy2021investigating, ye2022behavioral} focused on multi-step, process-oriented tasks that require sustained progress over time.
The study in \cite{martin2015effects,edwards2015examining}, tested email alerts for month-long programming assignments. Students received reminder emails seven, four, and two days before the deadline for two assignments, and an additional email ten days before for two assignments. The emails analyzed students' latest submissions and provided feedback on their progress relative to the course expectation and their peers. 
Similarly, \cite{brown2021nudging} studied \texttt{class-bot}, a GitHub bot integrated into each student’s assignment repository in an introductory programming course. Following a rubric with a checklist aligned with the software development cycle, the bot automatically evaluated the student's repository against the rubric and updated a GitHub issue daily to show pass/fail status for each checklist item.
Moreover, \cite{pereira2022struggling} developed \texttt{GanttBot}, a Telegram-based bot to support students on a three-month software engineering capstone project. The bot monitored progress against each student's Gantt chart, sent reminders as deadlines approached, and provided additional support such as task rescheduling and motivational messages.
In contrast, two studies \cite{bernuy2021investigating, ye2022behavioral} examined email reminders for short weekly exercises in large introductory CS courses. \citet{bernuy2021investigating} tested three different send times, between 48 hours and 24 hours before the deadline. \citet{ye2022behavioral} tested email reminders with different subject lines (prompt vs. statement), message lengths (short or long) and send times (70 or 30 hours before deadline). A prompt poses a question to elicit planning ("When did you plan to do ...?") whereas a statement delivers a direct reminder or instruction ("Remember to start early on ...").

\emph{Reminders improve submission timeliness and performance in multi-step, process-oriented tasks.}
In a project-based programming course, task-aware reminders embedded directly into students' workflows led to earlier starts, more iterative code development (i.e., greater code churn), and higher project scores \cite{brown2021nudging}.
Similarly, in month-long programming assignments, email alerts that analyzed students' recent submissions resulted in earlier first submissions, earlier completion, fewer late submissions, and higher grades among students who engaged earlier \cite{martin2015effects}.
Extending these findings to an even longer time horizon, a schedule-aware chatbot reduced overdue days relative to planned timelines in a three-month software engineering capstone project \cite{pereira2022struggling}. 
Overall, the evidence indicates that reminders function best as process supports that align with the structure and duration of complex assignments.

\emph{In weekly exercises, reminders mainly increase engagement rather than change timing or performance.}
Across studies of short, highly structured weekly exercises, reminder emails primarily affected whether students engaged with the work, rather than when or how they worked. In a large introductory course, generic email reminders increased the proportion of students who attempted the weekly homework but did not significantly affect start time, finish time, completion rates, or homework scores \cite{bernuy2021investigating}. Similarly, in \cite{ye2022behavioral}, reminder emails for weekly exercises increased completion rates but did not affect when students started or finished their work.
Taken together, these findings suggest that for short, routine tasks, reminders function primarily as participation prompts rather than tools for shaping work patterns or outcomes.

\emph{Students' perceptions of reminders often differ from their measured behavioural effects.} 
In \cite{martin2015effects}, even though email alerts led to improved submission timeliness, most students did not perceive them as useful: 55\% students reported that the emails were a waste of time and only 24\% said the alerts caused them to start earlier. 
In contrast, for weekly exercises in \cite{bernuy2021investigating}, students' perceptions were more positive than the measured effects: 56\% of the students described the reminder emails as helpful or motivating even though the reminders did not change start time, finish time, completion rates, or performance.
In \cite{pereira2022struggling}, students' perceptions aligned with measured outcomes only for specific reminder components. Email alerts and automatic rescheduling were rated as helpful and coincided with fewer overdue days, whereas motivational messages were rated less favourably and showed no behavioural effects. 
Collectively, these findings suggest that evaluating reminder interventions requires considering both student perceptions and objective behavioural outcomes, as the two may diverge.

%% file: results_6_others.tex
\subsection{Psychological and Social Interventions}

Psychological interventions aim to reduce procrastination by improving students' self-regulation skills, while social interventions reshape the social context for student work. Three papers examined such interventions in computing courses \cite{she2024clearmind, edwards2015examining, yeckehzaare2023reducing}. Two studies of psychological interventions explicitly targeted students' time management, motivation, or coping strategies \cite{she2024clearmind, edwards2015examining}. \citet{she2024clearmind} adapted ACT (Acceptance and Commitment Therapy) through a two-session, in-person workshop addressing the causes of procrastination and coping strategies for students in introductory data science courses. \citet{edwards2015examining} evaluated two lightweight strategies designed to promote self-regulation. Reflective writing assignments required students to write short reflections on their time management strategies. The schedule sheet intervention asked students to create and update project plans by breaking work into smaller tasks with intermediate deadlines and provided automated feedback flagging incomplete or unrealistic plans. A third study \cite{yeckehzaare2023reducing} examined a social intervention that altered how students interacted with peers in an independent study course by introducing optional peer presentations around student-generated questions.

\emph{Psychological interventions are more effective when they provide structured guidance rather than minimal scaffolding.}
Students who attended the ACT-based workshop reported significant reductions in procrastination and anxiety compared to a control group, consistent with findings from non-computing contexts \cite{she2024clearmind}. In contrast, neither reflective writing nor schedule sheet planning led to meaningful changes in student behaviour \cite{edwards2015examining}. Together, these results suggest that effective psychological interventions in computing courses may require direct instruction, facilitated practice, and explicit support.

\emph{Social accountability and peer interaction can reduce procrastination and promote sustained engagement in less structured learning environments.}
In \cite{yeckehzaare2023reducing}, students initially procrastinated by delaying question creation until shortly before progress report emails. Introducing optional weekly presentation sessions led students to generate significantly more questions, particularly on presentation days, and to distribute their work more evenly across the semester. The intervention also encouraged some students to adopt more advanced learner roles, including improving peers' questions, sharing resources, and coordinating presentation activities.

%% file: 4_discussion.tex
\section{Discussion of Cross-Category Patterns}
\label{sec:discussion}

Beyond summarizing patterns within individual intervention categories, this section synthesizes findings across categories to address RQ2 on how the interventions influence student behaviour and course outcomes. 
Across categories, the findings suggest that temporal structure, rather than incentive alone, serves as the primary driver of earlier student engagement. Effective interventions function by redistributing effort over time rather than increasing total workload. The shift toward earlier engagement and distributed work acts as a critical mediator for performance gains. However, the magnitude of these gains depends strongly on task structure, with greater benefits for long-horizon, multi-step assignments than for short, routine tasks. Furthermore, the results show a frequent divergence between student perceptions and measured behavioural and performance effects, indicating the need to evaluate both. Also, supportive designs consistently outperform punitive or restrictive schemes, aligning improved outcomes with positive student perceptions. Finally, uniform designs tend to benefit higher-performing students, indicating a need for personalization to better support those at risk of procrastination.

%% file: 5_limitations.tex
\section{Limitations}

First, this review does not make strong claims about procrastination research outside English-speaking contexts, as only studies published in English were included.
Second, the search strategy imposed constraints that may have limited coverage: reliance on specific keywords may have excluded studies using related terminology (e.g., time management or self-regulated learning), and the review was restricted to six major databases without extensive snowballing or manual reference searches. Broadening the search strategy would have substantially expanded the scope but reduced the feasibility of the review. Finally, although an iterative, consensus-based synthesis was used to mitigate bias, the categorization and interpretation of findings remain inherently subjective and may reflect the researchers’ perspectives.


%% file: 6_conclusion.tex
\section{Conclusion and Future Work}
\label{sec:conclusion}

A variety of interventions have been evaluated to reduce academic procrastination in post-secondary computing education. Evidence from 19 studies suggests that the most effective interventions reshape students’ work patterns over time. Across categories, structured and supportive designs encourage earlier engagement and distributed effort, with the strongest benefits observed for long, multi-step assignments. Interventions centered on penalties or incentives alone show less consistent effects.


Future work should examine psychological interventions that are proven in other fields but underexplored in computing  \cite{van2018overcoming, salguero2023interventions, rozental2018targeting}.
Second, research should investigate personalized interventions that account for differences in students’ performance levels and motivational profiles, focusing on supporting students with lower self-regulation skills. 
Finally, future studies should identify which forms of structure, feedback, or nudging are effective for short, routine tasks.